\documentclass[12pt]{iopart}

\usepackage{iopams, upgreek, amssymb, graphicx, subfigure, paralist, dsfont, dcolumn, subfigure}
\usepackage[colorlinks, linkcolor=blue, citecolor=blue, urlcolor=blue, breaklinks=true]{hyperref}

\renewcommand{\eref}[1]{Eq.~(\ref{#1})}
\renewcommand{\fref}[1]{Fig.~\ref{#1}}
\newcommand{\aref}[1]{\ref{#1}}
\renewcommand{\sref}[1]{Sec.~\ref{#1}}
\renewcommand{\tref}[1]{Table~\ref{#1}}

\newcommand{\eg}{\emph{e.g.}}

\newcommand{\abs}[1]{\left|#1\right|}

\newcommand{\re}{\mathrm{Re}}

\def\Oin{\Omega_{\mathrm{in},\lambda}}

\def\Oc{\Omega_{\mathrm{c},\lambda}}

\def\Oinnolambda{\Omega_\mathrm{in}}
\def\Ocnolambda{\Omega_\mathrm{c}}
\newcommand\hc{\mathrm{H.c.}}

\makeatletter
\newlength \figurewidth
\makeatother

\begin{document}

\title{Exciton-mediated photothermal cooling in GaAs membranes}
\author{Andr\'e Xuereb$^1$, Koji Usami$^2$, Andreas Naesby$^2$, Eugene S.\ Polzik$^2$, and Klemens Hammerer$^3$}
\address{$^1$\ Centre for Theoretical Atomic, Molecular and Optical Physics, School of Mathematics and Physics, Queen's University Belfast, Belfast BT7\,1NN, United Kingdom}
\address{$^2$\ QUANTOP -- Danish National Research Foundation Center for Quantum Optics, Niels Bohr Institute, Blegdamsvej 17, 2100 Copenhagen, Denmark}
\address{$^3$\ Institut f\"ur Gravitationsphysik, Leibniz Universit\"at Hannover, Callinstra\ss{}e 38, D-30167 Hannover, Germany; and Institut f\"ur Theoretische Physik, Leibniz Universit\"at Hannover, Appelstra\ss{}e 2, D-30167 Hannover, Germany}
\eads{\mailto{andre.xuereb@qub.ac.uk}, \mailto{usami@nbi.dk}, \mailto{naesby@nbi.dk}, \mailto{polzik@nbi.dk}, \mailto{klemens.hammerer@itp.uni-hannover.de}}

\date{\today}

\begin{abstract}
Cooling of the mechanical motion of a GaAs nano-membrane using the photothermal effect mediated by excitons was recently demonstrated by some of us [K.\ Usami, \emph{et al.}, Nature Phys.\ \textbf{8}, 168 (2012)] and provides a clear example of the use of thermal forces to cool down mechanical motion. Here, we report on a single-free-parameter theoretical model to explain the results of this experiment which matches the experimental data remarkably well.
\end{abstract}

\maketitle

\begin{figure}[t]
 \centering
 \includegraphics{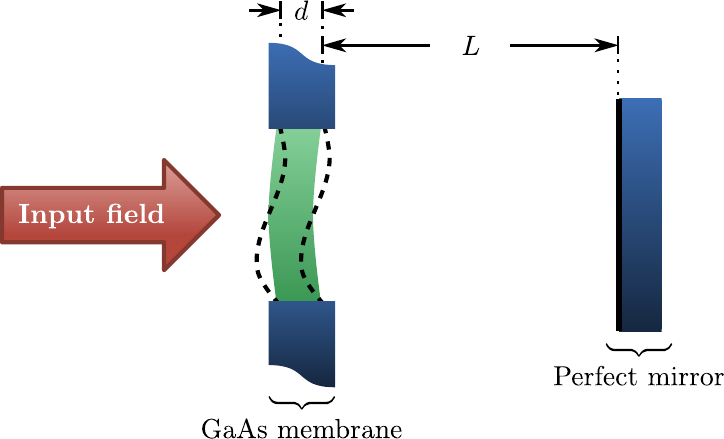}
 \caption{Schematic of the experiment. A Fabry--P\'erot cavity is bounded by a perfect mirror on one end and a GaAs membrane on the other. Drumhead vibrations of the membrane, indicated by the dashed lines, couple to the cavity field and to the excitons inside the membrane. In the experiment, $L=2.9$\,cm, $d=160$\,nm, and the membrane is not perfectly flat; this latter feature is important to the mechanism we consider, as explained in \sref{sec:OMC}.}
 \label{fig:System}
\end{figure}
\section{Introduction and motivation}
Micro- and nanomechanical systems in the quantum regime offer exciting perspectives for fundamental tests of quantum physics as well as for quantum technological applications. It is a unique feature of these systems that they can be strongly coupled to a plethora of other quantum systems:\ optomechanics explores interactions with light based on radiation pressure~\cite{Kippenberg2008,Aspelmeyer2010}, dipole gradient~\cite{Li2009,Xuereb2012a}, or photothermal forces~\cite{Pinard2008,DeLiberato2011,Restrepo2011}; electromechanics investigates the coupling of electronic and mechanical degrees of freedom~\cite{OConnell2010,Taylor2011}; and magnetic forces can couple a mechanical oscillator to magnetic moments~\cite{Forstner2012}, even of single electrons~\cite{Rugar2004}. It is this versatile nature of mechanical systems which makes them attractive as basic building blocks for hybrid quantum systems~\cite{Wallquist2009b}. In the current paper we elaborate on the recent findings reported in Ref.\ \cite{Usami2012} demonstrating the interplay of photonic and electronic degrees of freedom in a micromechanical semiconductor membrane.\par
The photothermal effect~\cite{Pinard2008,DeLiberato2011,Restrepo2011,Metzger2004,Metzger2008} elegantly overcomes one fundamental limit encountered by any optomechanical cooling mechanism based on the radiation-pressure interaction~\cite{Kippenberg2008,Aspelmeyer2010}. That is to say, a single photon of frequency $\omega_\mathrm{L}$ can only provide an energy change $\Delta E\sim(v/c)\hbar\omega_\mathrm{L}$ upon reflection off a mirror with velocity $v$, and this due to the Doppler shift~\cite{Restrepo2011}, but absorption of the same photon, as per the photothermal effect, implies $\Delta E\sim\hbar\omega_\mathrm{L}$. How this $\Delta E$ translates to a change in motional energy is a less well-defined concept, but it is clear that, in principle, the latter effect can give rise to cooling forces that eclipse the radiation-pressure force produced by the same number of photons.\par
In the semiconductor GaAs, the absorption of an above-bandgap optical photon and subsequent decay of the associated bound states~\cite{Segall1968,Gartner1961} is a complicated process; the energy liberated by the photon first creates an electron--hole bound pair (an exciton), which decays by scattering phonons throughout the structure of the material. This scattering process, which manifests itself primarily as the transport of heat and takes place over the \emph{thermalisation time} $\tau_\mathrm{th}$, changes the properties of the material. The drumhead modes of a membrane are critically dependent on these properties, and this process therefore couples the absorption of the light to the motion of these modes. Our main aim in this paper is to describe a phenomenological Hamiltonian model for this process and to predict the cooling or heating effect imparted by this photothermal interaction. We take particular care to model the details of the experimental system of Ref.~\cite{Usami2012}; we notice, for example, that the coupling of the excitonic field to the continuum of input field modes is crucial to describing the experimental data.\\
This paper is structured as follows. In the next section we shall build our system Hamiltonian and derive the equations of motion of the field operators. Following this, we shall insert a memory kernel in these equations to account for the delay in the thermalisation process. After we fit our model to the experimental data, we briefly discuss the possibility of using pure deformation-potential effects to achieve cooling in similar systems, and then conclude.

\section{System Hamiltonian and Equations of motion}\label{sec:HamEoM}
\begin{figure*}[t]
 \centering
  \includegraphics[scale=1.2]{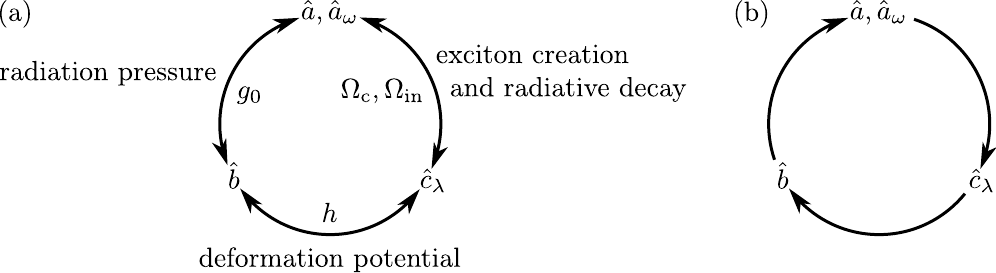}
 \caption{(a)~The full set of interactions that our model describes; each interaction is labelled by the relevant coupling constant in the effective description of \sref{sec:AEMem}. (b)~Adiabatic elimination allows us to describe the system in terms of one cycle of interactions, with the mechanics modifying the optical fields, which act on the excitonic fields, which in turn couple to the mechanics. A memory kernel, used to describe a delayed interaction, will later be introduced into the segment connecting $\hat{c}_\lambda$ and $\hat{b}$.}
 \label{fig:Processes}
\end{figure*}
Let us start by considering the physical system represented in \fref{fig:System}:\ a Fabry--P\'erot cavity is bounded by a perfect mirror on one end and a semi-transmissive GaAs membrane on the other. We shall describe the exciton fields inside the membrane by means of a bosonic approximation, which is valid when the exciton population is not too large, and assign to these fields the operators $\hat{c}_\lambda$ (frequency $\omega_\lambda$ and decoherence rate $\gamma_\lambda$), where $\lambda$ is some, possible continuous, index labelling the modes. Any sum over $\lambda$ is to be interpreted as either a sum, if $\lambda$ is discrete, or an integral, otherwise.\par
These fields interact with the cavity field, $\hat{a}$ (frequency $\omega_\mathrm{c}$ and HWHM linewidth $\kappa_\mathrm{c}$), as well as with an infinity of field modes $\hat{a}_\omega$ that represent the free field that forms the input to the cavity field. The excitonic fields also interact with our chosen mechanical mode, which we describe using the operator $\hat{b}$, and which has a mechanical frequency $\omega_\mathrm{m}$ and HWHM linewidth $\kappa_\mathrm{m}$.\par
\begin{figure*}[t]
 \centering
 \subfigure[\ Dataset 1; $\lambda_\mathrm{L}=870$\,nm]{%
  \includegraphics[width=\figurewidth]{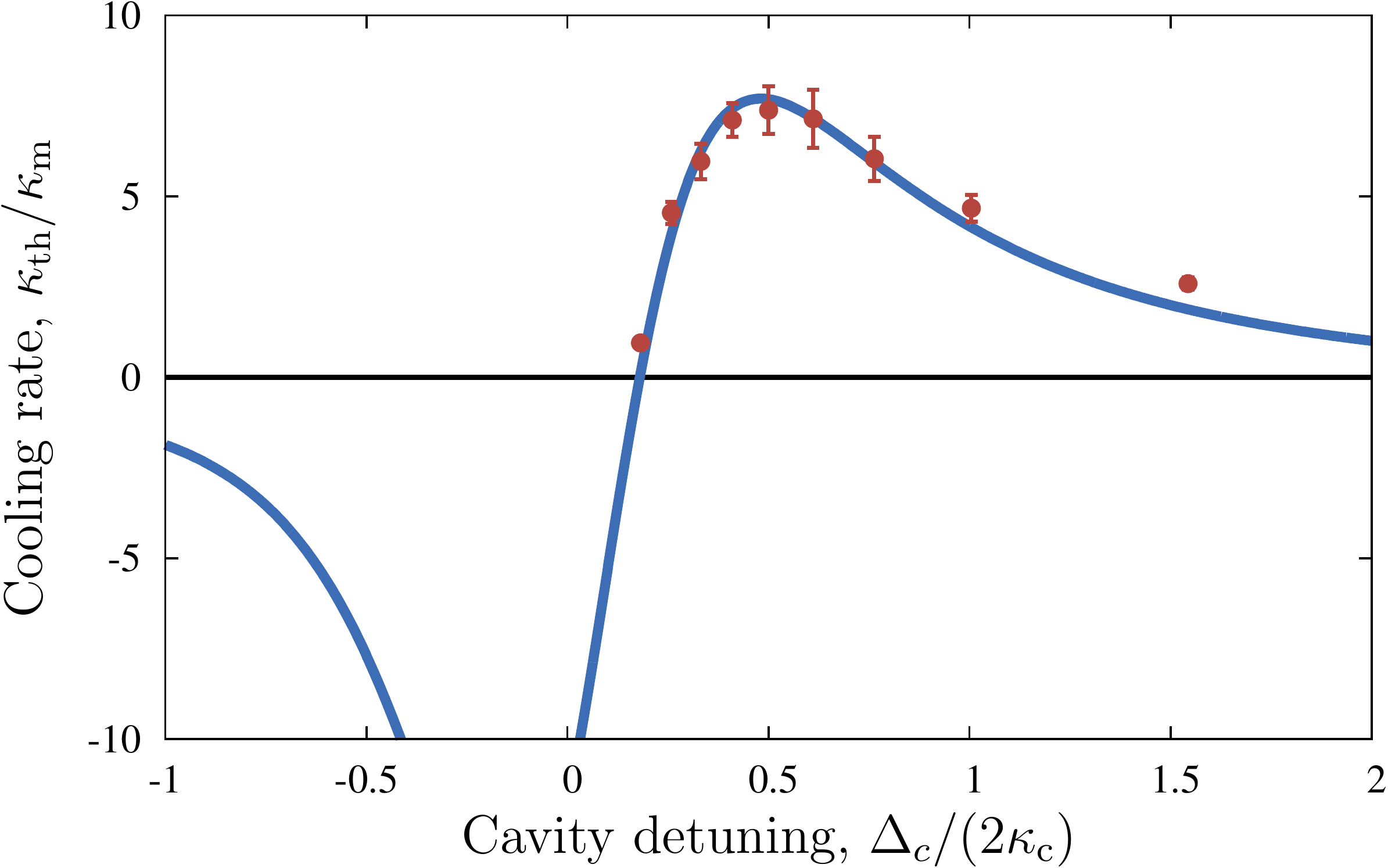}}\quad
 \subfigure[\ Dataset 2; $\lambda_\mathrm{L}=852$\,nm]{%
 \includegraphics[width=\figurewidth]{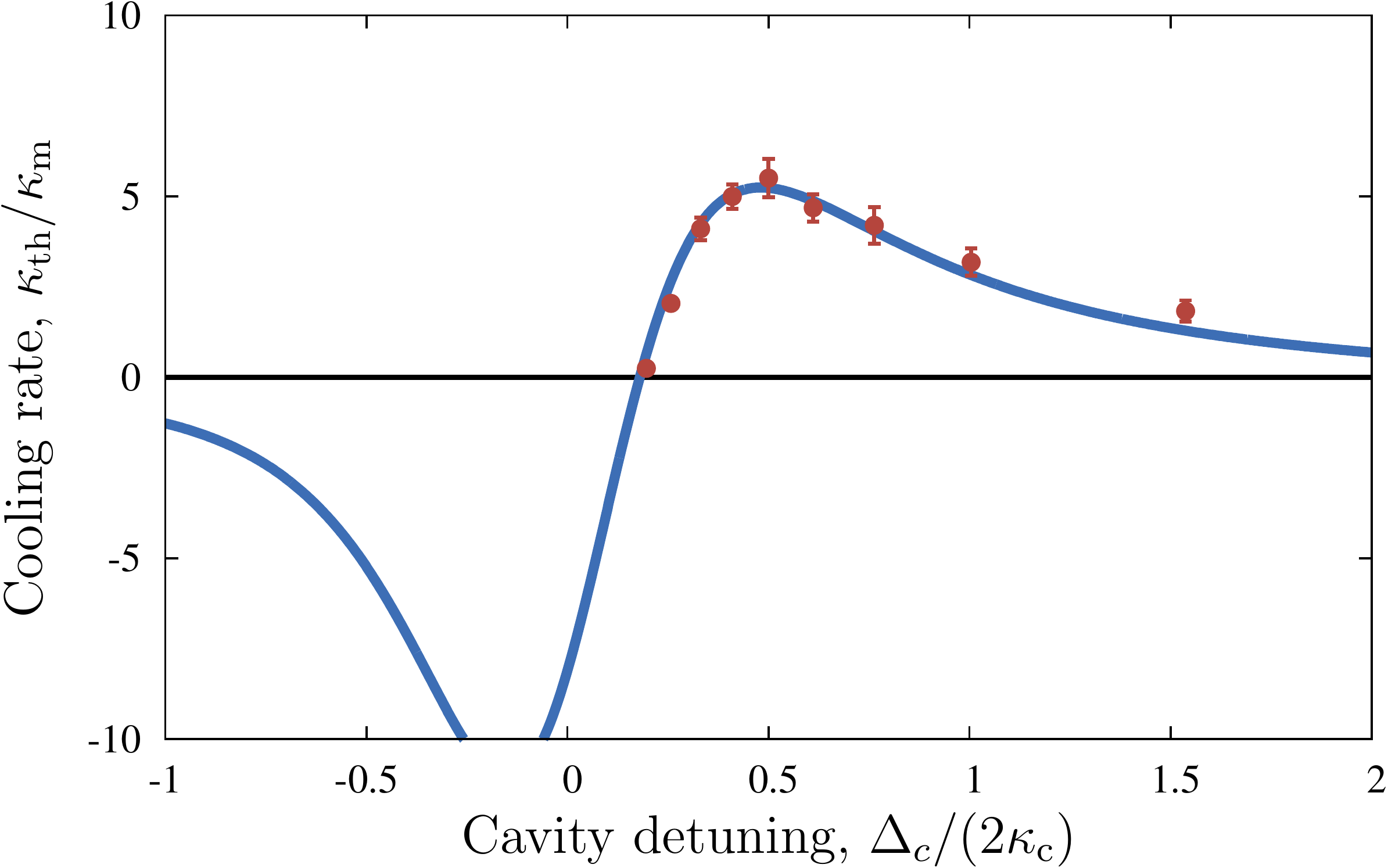}}\\
 \subfigure[\ Dataset 3; $\lambda_\mathrm{L}=852$\,nm]{%
 \includegraphics[width=\figurewidth]{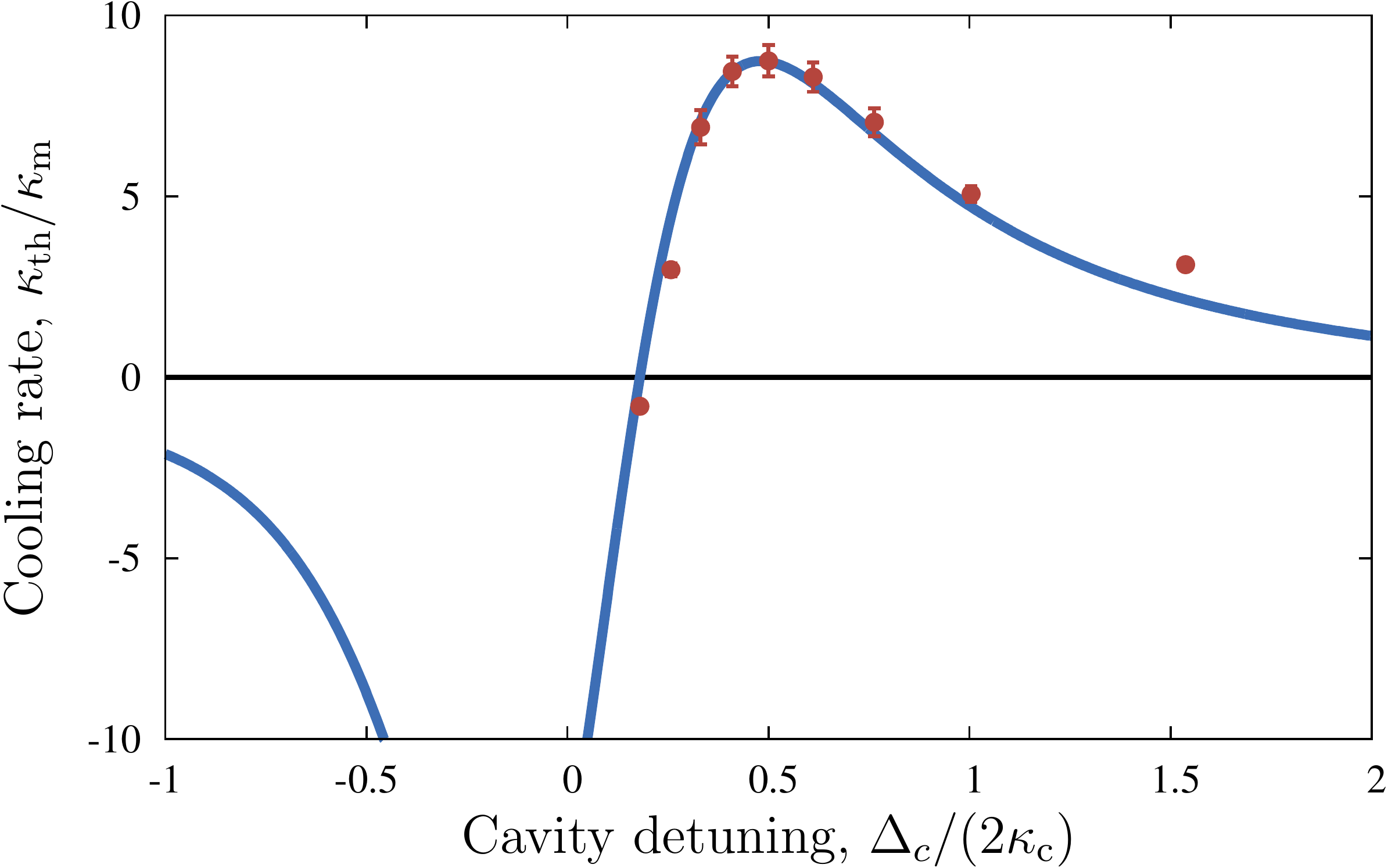}}\quad
 \subfigure[\ Dataset 4; $\lambda_\mathrm{L}=852$\,nm]{%
 \includegraphics[width=\figurewidth]{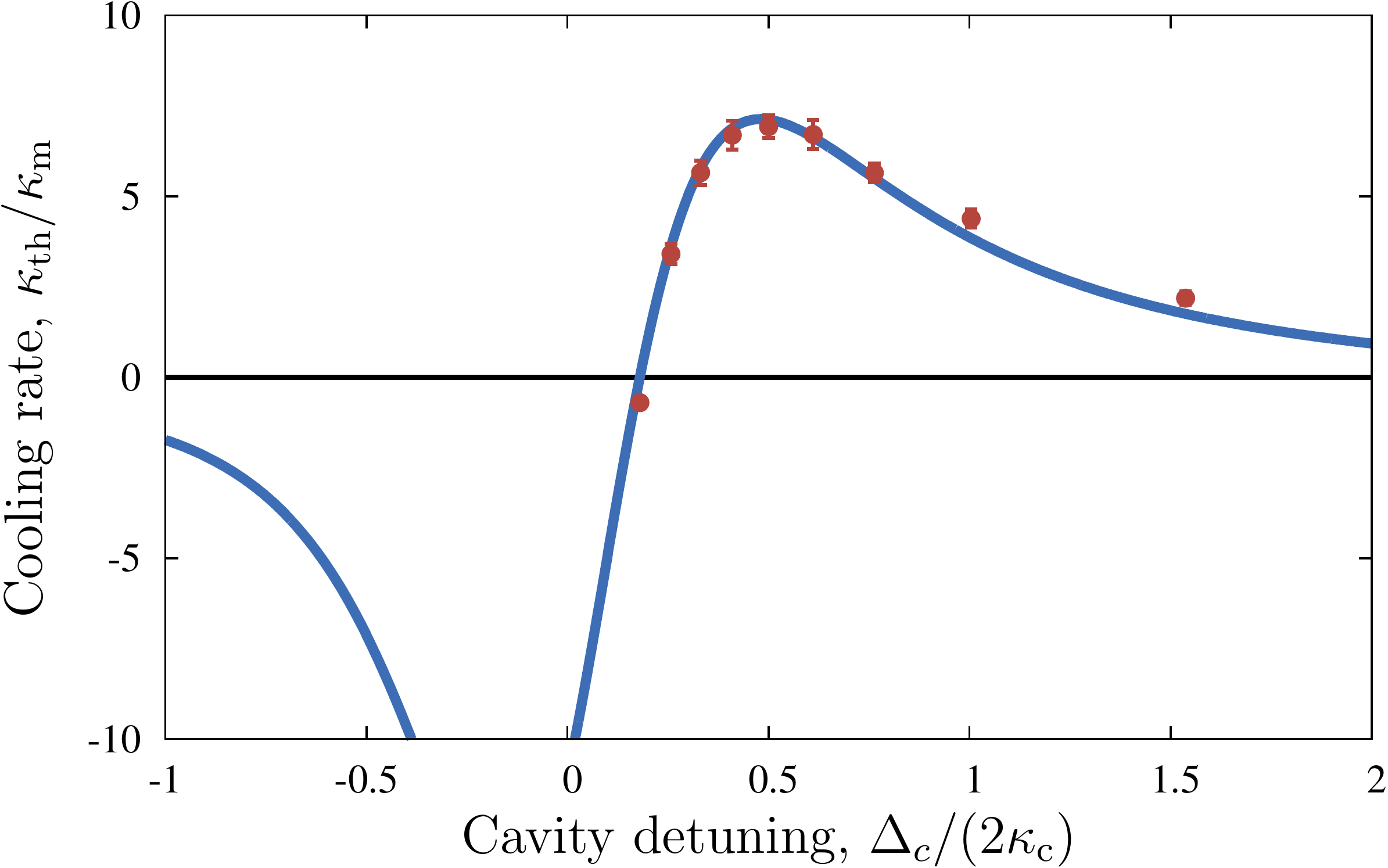}}
 \caption{Fits to four sets of experimental data; Dataset 1 is the set reported in Ref.~\cite{Usami2012}. The four sets of data differed in the location of the driving beam on the membrane and in the driving wavelength $\lambda_\mathrm{L}$ used. All other system parameters are reported in the text and in \tref{tab:FitParams}.}
 \label{fig:Fits}
\end{figure*}

The Hamiltonian $\hat{H}$ is made up of four different contributions. The free Hamiltonian reads (we shall take $\hbar=1$ throughout this paper for conciseness of notation)
\begin{equation}
\hat{H}_\mathrm{free} = \omega_c\hat{a}^\dagger \hat{a}
+\int\rmd\omega\,\omega\hat{a}_\omega^\dagger\hat{a}_\omega+\sum_\lambda\omega_\lambda\hat{c}_\lambda^\dagger\hat{c}_\lambda+\omega_\mathrm{m}\hat{b}^\dagger\hat{b}\,,
\end{equation}
and describes the fields in the absence of any interaction or dissipation. The next set of terms describes dissipation, and reads
\begin{equation}
\hat{H}_\mathrm{diss} = i\sqrt{\frac{\kappa_\mathrm{c}}{\pi}}\int\rmd\omega\,\bigl(\hat{a}_\omega^\dagger\hat{a}-\hat{a}_\omega\hat{a}^\dagger\bigr)+\hat{H}_\mathrm{mech,diss}+\hat{H}_\mathrm{exc,diss}\,;
\end{equation}
we have chosen to write down explicitly only the Hamiltonian describing the cavity field decay. $\hat{H}_\mathrm{mech,diss}$ ($\hat{H}_\mathrm{exc,diss}$) similarly describes the dissipation of the mechanical (excitonic) operator(s). There are two sets of interaction terms, illustrated pictorially in \fref{fig:Processes}(a); the first couples the excitons to the electric field:
\begin{equation}
\label{eq:Habs}
\hat{H}_\mathrm{abs} = \sum_\lambda\Biggl(\int\rmd\omega\,\frac{\Oin}{\sqrt{\pi\kappa_\mathrm{c}}}\hat{a}_\omega+\Oc\hat{a}\Biggr)\hat{c}_\lambda^\dagger+\hc\,,
\end{equation}
where $\Oc$, which is assumed to be real, is the coupling rate of exciton mode $\lambda$ with the cavity field, and $\Oin$ with the free-field modes. The coupling of the exciton modes to the free field cannot be neglected in this case:\ the membrane is thick enough for interference effects between the input field and the cavity field to be significant, and $\abs{\Oin/\Oc}$ to be of order unity. The two terms in $\hat{H}_\mathrm{abs}$ therefore interfere, leading to an asymmetric cooling spectrum (see \sref{sec:OMC}, below, and \fref{fig:Fits}); this phenomenon is closely linked to the Fano line-shapes observed in optomechanical systems where the mechanical oscillator is coupled both to a cavity field and to the free field~\cite{Elste2009,Xuereb2011b}, although the dominant coupling of the optical fields to the mechanics is indirect in the present case.\\
The second set of interaction terms describes the coupling of the mechanical motion to the cavity field and to the excitons~\cite{Toyozawa1958,SteynRoss1983}:
\begin{equation}
\hat{H}_\mathrm{mech} = g_0\hat{a}^\dagger\hat{a}\bigl(\hat{b}+\hat{b}^\dagger\bigr)+\sum_{\lambda,\lambda^\prime}h_{\lambda,\lambda^\prime}\hat{c}_\lambda^\dagger\hat{c}_{\lambda^\prime}\bigl(\hat{b}+\hat{b}^\dagger\bigr)\,.
\end{equation}
$g_0$ is the usual (radiation-pressure) coupling constant and $h_{\lambda,\lambda^\prime}$ describes the deformation-potential coupling. We omit any direct coupling of the motion to the free field, since such effects would be very small~\cite{Xuereb2011b} compared to the terms in the preceding equation. Finally, we can write
\begin{equation}
\hat{H} = \hat{H}_\mathrm{free} + \hat{H}_\mathrm{diss} + \hat{H}_\mathrm{abs} + \hat{H}_\mathrm{mech}\,.
\end{equation}
This Hamiltonian can be used to generate the Heisenberg--Langevin equations of motion for the field operators. We choose to work in a frame rotating at the frequency of the driving field, $\omega_\mathrm{L}$, and define the detunings $\Delta_\mathrm{c}=\omega_\mathrm{L}-\omega_{c}$ and $\Delta_\lambda=\omega_\mathrm{L}-\omega_\lambda$. At this point we shall make two further assumptions regarding the exciton fields. The quantities $\Oc\to\Ocnolambda$, $\Oin\to\Oinnolambda$, $h_{\lambda,\lambda^\prime}\to h_0$, and $\gamma_\lambda\to\gamma$ are assumed to be independent of the index $\lambda$. We also assume that the exciton density of states is constant in the relevant region, which is a good approximation for pumping well above the band-gap energy. In order to avoid introducing new symbols we shall displace each operator $\hat{o}$ by its mean value $\bar{o}$:\ $\hat{o}\to\bar{o}+\hat{o}$, where the operator on the right-hand side has zero mean. Therefore, \emph{all operators in the following will have zero mean}. The linearised equations of motion can now be written as
\begin{eqnarray}
\dot{\hat{a}}=-(\kappa_\mathrm{c}-i\Delta_\mathrm{c})\hat{a}-\sqrt{2\kappa_\mathrm{c}}\hat{a}_\mathrm{in}-ig_0\bar{a}\bigl(\hat{b}+\hat{b}^\dagger\bigr)\,,\\
\dot{\hat{b}}=-(\kappa_\mathrm{m}+i\omega_\mathrm{m})\hat{b}-\sqrt{2\kappa_\mathrm{m}}\hat{b}_\mathrm{in}-ig_0\bigl(\bar{a}^\star\hat{a}+\bar{a}\hat{a}^\dagger\bigr)-i\sum_\lambda\bigl(h^\star\hat{c}_\lambda+h\hat{c}_\lambda^\dagger\bigr)
\end{eqnarray}
and
\begin{eqnarray}
\dot{\hat{c}}_\lambda=-(\gamma-i\Delta_\lambda)\hat{c}_\lambda-\sqrt{2\gamma}\hat{c}_{\lambda,\mathrm{in}}-ih\bigl(\hat{b}+\hat{b}^\dagger\bigr)-i\bigl(\Oinnolambda+\Ocnolambda\bigr)\hat{a}\nonumber\\\qquad\qquad-i\sqrt{\frac{2}{\kappa_\mathrm{c}}}\Oinnolambda\hat{a}_\mathrm{in}\,,
\end{eqnarray}
where terms without a significant contribution were dropped, and where we defined $h=h_0\sum_\lambda\bar{c}_\lambda$.. The operators $\hat{c}_{\lambda,\mathrm{in}}$ describe the zero-mean Langevin forces associated with the excitonic modes, whereas the optical field input operator $\hat{a}_\mathrm{in}$ is defined as per the usual input--output theory (cf.\ Ref.~\cite[\S 5.3]{Gardiner2004}), but we note that the interaction of the excitons with the free field modifies the input--output relation for the system, yielding
\begin{equation}
\hat{a}_\mathrm{out}=\hat{a}_\mathrm{in}+\sqrt{2\kappa_\mathrm{c}}\hat{a}-i\sqrt{\frac{2}{\kappa_\mathrm{c}}}\Oinnolambda^\star\sum_\lambda\hat{c}_\lambda\,.
\end{equation}
The mean values of the fields satisfy ($\bar{c}_\mathrm{in}=0$)
\begin{eqnarray}
-(\kappa_\mathrm{c}-i\Delta_\mathrm{c})\bar{a}-i\bigl(\Ocnolambda-\Oinnolambda\bigr)^\star\sum_\lambda\bar{c}_\lambda-\sqrt{2\kappa_\mathrm{c}}\bar{a}_\mathrm{in}=0\,,\mathrm{ and}\\
-(\gamma-i\Delta_\lambda)\bar{c}_\lambda-i\bigl(\Oinnolambda+\Ocnolambda\bigr)\bar{a}-i\sqrt{\frac{2}{\kappa_\mathrm{c}}}\Oinnolambda\bar{a}_\mathrm{in}=0\,.
\end{eqnarray}
We have absorbed $\bar{b}$ into an effective redefinition of $\Delta_\mathrm{c}$ and $\Delta_\lambda$, and therefore we have $\bar{b}=0$. It is now apparent that the equations for the optical and mechanical fields involve only sums of the type $\sum_\lambda\hat{c}_\lambda$ or $\sum_\lambda\bar{c}_\lambda$. These sums can be performed easily due to our assumptions, yielding
\begin{equation}
\bar{a}=\Biggl(-\sqrt{\frac{2}{\kappa_\mathrm{c}}}\bar{a}_\mathrm{in}\Biggr)\frac{\kappa_\mathrm{c}\gamma+\nu\Oinnolambda\bigl(\Ocnolambda-\Oinnolambda\bigr)^\star}{\bigl(\kappa_\mathrm{c}-i\Delta_\mathrm{c}\bigr)\gamma+\nu\bigl(\Ocnolambda+\Oinnolambda\bigr)\bigl(\Ocnolambda-\Oinnolambda\bigr)^\star}\,,
\end{equation}
and
\begin{equation}
\sum_\lambda\bar{c}_\lambda=\sqrt{\nu}\Biggl(\sqrt{\frac{2}{\kappa_\mathrm{c}}}\bar{a}_\mathrm{in}\Biggr)\frac{i\kappa_\mathrm{c}\sqrt{\nu}\Ocnolambda-\Delta_\mathrm{c}\sqrt{\nu}\Oinnolambda}{\bigl(\kappa_\mathrm{c}-i\Delta_\mathrm{c}\bigr)\gamma+\nu\bigl(\Ocnolambda+\Oinnolambda\bigr)\bigl(\Ocnolambda-\Oinnolambda\bigr)^\star}\,,
\end{equation}
where $\nu\equiv\gamma\sum_\lambda\bigl(\gamma-i\Delta_\lambda\bigr)^{-1}\in\mathbb{R}^+$ accounts for the number of exciton modes we are interacting with.

\section{Adiabatic elimination and memory kernel}\label{sec:AEMem}
The system we aim to describe has a hierarchy of dynamics determined by $\gamma\gg\kappa_\mathrm{c}\gg\kappa_\mathrm{m},\omega_\mathrm{m}$. Sequential adiabatic elimination of the fields is therefore possible, first solving the equation of motion for the excitonic fields, on whose time-scale $\hat{a}$ is approximately constant, and then for the cavity field. We concentrate on the regime where the exciton-mediated effect dominates over the radiation pressure force. At this level of approximation we can describe the web of interactions in our system via the following process:
\begin{enumerate}
\item the position of the mechanical oscillator changes the photon number inside the cavity ($\hat{b}$ affects $\hat{a})$,
\item the exciton populations follow the changing cavity field ($\hat{a}$ affects $\hat{c}_\lambda$), and finally
\item the changing exciton populations modulate the mechanical properties, coupling to $\hat{b}$ ($\hat{c}_\lambda$ affects $\hat{b}$).
\end{enumerate}
\fref{fig:Processes} illustrates the different interactions that take place in the system we are describing, together with the reduced system that results after adiabatically eliminating the optical and excitonic fields.\\
With the above in mind, and ignoring the contributions from the input field, we can write the adiabatic solution of the cavity field operator familiar from the radiation-pressure cooling literature~\cite{WilsonRae2007,Marquardt2007}:
\begin{equation}
\hat{a}\approx-ig_0\bar{a}\Biggl[\frac{1}{\kappa_\mathrm{c}-i(\Delta_\mathrm{c}+\omega_\mathrm{m})}\hat{b}+\frac{1}{\kappa_\mathrm{c}-i(\Delta_\mathrm{c}-\omega_\mathrm{m})}\hat{b}^\dagger\Biggr]\,,
\end{equation}
which is substituted into the equation of motion for $\hat{c}_\lambda$ to give
\begin{equation}
\hat{c}_\lambda\approx-\frac{g_0\bar{a}\bigl(\Ocnolambda+\Oinnolambda\bigr)}{\gamma-\kappa_\mathrm{c}-i(\Delta_\lambda-\Delta_\mathrm{c})}\Biggl[\frac{1}{\kappa_\mathrm{c}-i(\Delta_\mathrm{c}+\omega_\mathrm{m})}\hat{b}+\frac{1}{\kappa_\mathrm{c}-i(\Delta_\mathrm{c}-\omega_\mathrm{m})}\hat{b}^\dagger\Biggr]\,.
\end{equation}
We see from this equation that the function of the excitons at this level of approximation is to act as a channel for the absorbed optical energy to interact with the mechanics. Noting that $\gamma$ is much larger than all the other frequencies, we obtain
\begin{eqnarray}
\sum_\lambda\hat{c}_\lambda\approx-\frac{\nu}{\sqrt{\nu}}\frac{g_0\bar{a}\sqrt{\nu}\bigl(\Ocnolambda+\Oinnolambda\bigr)}{\gamma}\Biggl[&\frac{1}{\kappa_\mathrm{c}-i(\Delta_\mathrm{c}+\omega_\mathrm{m})}\hat{b}\nonumber\\
&\qquad\quad+\frac{1}{\kappa_\mathrm{c}-i(\Delta_\mathrm{c}-\omega_\mathrm{m})}\hat{b}^\dagger\Biggr]\,.
\end{eqnarray}
Substitution of this solution into the equation of motion for $\hat{b}$ gives us the adiabatic dynamics when the excitons couple to the mechanics through the deformation potential. Because we want to describe the \emph{time-delayed} effect of the excitons on the mechanics, however, we must introduce a memory kernel $\mathcal{M}(t)$ into the equation of motion for $\hat{b}$. This follows the ideas outlined in related treatments of the photothermal effect~\cite{Metzger2008,Pinard2008,DeLiberato2011,Restrepo2011}, and in our notation corresponds to setting:
\begin{equation}
\label{eq:IntroMK}
h_0^\star\sum_\lambda\hat{c}_\lambda\rightarrow\eta^\star\int_{-\infty}^\infty\mathcal{M}(t-\tau)\sum_\lambda\hat{c}_\lambda(\tau)\,\rmd\tau\,,
\end{equation}
and similarly for the term involving $h_0$. This process changes the physical meaning of these terms. We emphasise that $\eta$ no longer describes the deformation-potential coupling, but is a phenomenological coupling constant that describes the strength of the delayed interaction linking the exciton fields with the mechanical motion. Physically, \eref{eq:IntroMK} tells us that the modulation to the mechanical properties takes into account the entire history of the exciton fields. We choose to use an exponentially-decaying memory kernel~\cite{Restrepo2011}:
\begin{equation}
\mathcal{M}(t-\tau)=\frac{1}{\tau_\mathrm{th}}e^{-(t-\tau)/\tau_\mathrm{th}}\Theta(t-\tau)\,,
\end{equation}
where $\Theta(t)$ is the Heaviside step function and accounts for the causal nature of the memory kernel; this $\mathcal{M}(t)$ leads to the same expressions as the ``$h(t)$'' chosen by Metzger and co-workers, cf.\ Ref.~\cite{Metzger2008}, upon integration by parts of the relevant terms. The use of a memory kernel in the equation of motion for $\hat{b}$ can be motivated by making use of an extended model that includes a bath of phonon modes which act as the intermediary between the excitons and $\hat{b}$. Elimination of the these modes in $\rmd\hat{b}/\rmd t$ naturally gives rise to a time-integral of a sum of decaying exponentials, which we identify with an exponentially-decaying memory kernel; this process is outlined in \aref{app:MemoryKernel}.\par
A brief note about the effect of noise terms is due. It lies outside the scope of this paper to consider the effects of noise on the limits of this cooling mechanism; being interested in cooling rates in this article, we accordingly discard such terms. As discussed in Ref.~\cite{Restrepo2011}, the nature of the photothermal effect does not preclude reaching the ground state, even in the bad-cavity limit, despite the absorption of light in the mechanical oscillator. The thermal noise induced by the absorbed light can be effectively modelled as a Langevin force term as in, \eg, Eq.~(5) of Ref.~\cite{Pinard2008} or Eq.~(15) of Ref.~\cite{Restrepo2011}. This Langevin force can be viewed as having its physical origins in the thermal fluctuations of the phononic bath described in \aref{app:MemoryKernel}, which couples the exciton modes to $\hat{b}$.\\
A unique feature of our setup lies in the interference between the cavity and input fields, discussed after \eref{eq:Habs} above, which could, in analogy with dissipative optomechanics~\cite{Xuereb2011b}, lead to a situation where the effect of the noise originating from the optical fields cancels out, and therefore to a more efficient cooling mechanism and a lower base temperature.

\begin{table*}
\caption{\label{tab:FitParams}Experimental parameters and coupling constant extracted from the data. The datasets are numbered as per \fref{fig:Fits}.}
\footnotesize
\centering
\begin{tabular}{c|ccccc|c}
\br
 & $\lambda_\mathrm{L}$ [nm] & $P_\mathrm{in}$ [$\upmu$W] & $\kappa_\mathrm{m}$ [s$^{-1}$] & $f_\mathrm{abs}$ [\%] & $\Ocnolambda^2/\gamma$ [$2\pi$\,MHz] & $\eta_\mathrm{th}/\gamma\times10^2$\\
\hline
$1$ & $870$ & $20$ & $1.8$ & $50$ & $32.3$ & 7.5\\
$2$ & $852$ & $25$ & $2.2$ & $55$ & $35.4$ & 4.6\\
$3$ & $852$ & $25$ & $2.2$ & $55$ & $35.4$ & 7.6\\
$4$ & $852$ & $25$ & $2.2$ & $55$ & $35.4$ & 6.2\\
\br
\end{tabular}
\end{table*}

\section{Optomechanical cooling rate}\label{sec:OMC}
Proceeding from the previous section along the same lines as standard optomechanical theory, we can now derive a simple expression for the optomechanical cooling rate due to this photothermal effect. Indeed, we can show that the mechanical decay rate changes from $\kappa_\mathrm{m}$ to $\kappa_\mathrm{m}+\kappa_\mathrm{th}$, where for $\tau_\mathrm{th}\gg1/\omega_\mathrm{m}$ and $\Ocnolambda\in\mathbb{R}$
\begin{eqnarray}
\label{eq:kth}
\fl\kappa_\mathrm{th}=\frac{P_\mathrm{in}}{\hbar\omega_\mathrm{L}}\frac{2g_0}{\bigl(\kappa_\mathrm{c}^2+\Delta_\mathrm{c}^2\bigr)\omega_\mathrm{m}\tau_\mathrm{th}}\frac{\eta_\mathrm{th}}{\gamma}\frac{\Ocnolambda^2}{\gamma}\,&\re\Biggl\{\frac{\bigl(1+\Oinnolambda/\Ocnolambda\bigr)\bigl(\Delta_\mathrm{c}\Oinnolambda^\star/\Ocnolambda+i\kappa_\mathrm{c}\bigr)}{\kappa_\mathrm{c}-i(\omega_\mathrm{m}+\Delta_\mathrm{c})}\nonumber\\
\fl&\qquad\qquad+\frac{\bigl(1+\Oinnolambda^\star/\Ocnolambda\bigr)\bigl(\Delta_\mathrm{c}\Oinnolambda/\Ocnolambda-i\kappa_\mathrm{c}\bigr)}{\kappa_\mathrm{c}-i(\omega_\mathrm{m}-\Delta_\mathrm{c})}\Biggr\}\,,
\end{eqnarray}
with $P_\mathrm{in}=\hbar\omega_\mathrm{L}\abs{\bar{a}_\mathrm{in}}^2$ being the input power coupled into the cavity (in watts), the sum over $\lambda$ was absorbed into the phenomenological coupling constant $\eta_\mathrm{th}=\nu\eta$, and we have absorbed $\sqrt{\nu}$ into each of $\Ocnolambda$ and $\Oinnolambda$. In accordance with the approximations made during adiabatic elimination, \eref{eq:kth} excludes higher-order terms in $\Ocnolambda$ and $\Oinnolambda$. It is worth noting that the sign and magnitude of $\eta_\mathrm{th}$ depend on the shape of the membrane. The membrane in the experiment has a slight curvature, such that any thermal expansion has a well-defined effect on its effective position $\hat{x}\equiv\bigl(\hat{b}+\hat{b}^\dagger\bigr)/\sqrt{2}$; if the membrane were to be flipped over, the sign of $\eta_\mathrm{th}$ would change. For a perfectly flat membrane, the membrane `would not know' which way to buckle under thermal expansion; $\eta_\mathrm{th}$ would then be zero and other terms would be expected to dominate.\par
The expression for $\kappa_\mathrm{th}$ depends critically on $\Oinnolambda$, whose relationship to $\Ocnolambda$ is fixed by the geometry of the cavity and membrane. In the good-cavity limit, which is valid whenever the finesse of the cavity is $\gg1$, and taking into account the large refractive index of GaAs, we obtain
\begin{equation}
\frac{\Oinnolambda}{\Ocnolambda}=-\frac{i}{\sqrt{2}}e^{i(k_\mathrm{L}d/2-2L\Delta_\mathrm{c}/c)}\sin(k_\mathrm{L}d/2)\,,
\end{equation}
where $d$ is the thickness of the membrane, $L$ the length of the cavity, and $k_\mathrm{L}=\omega_\mathrm{L}/c$. $\Ocnolambda$ itself can be fixed by observing the fraction $f_\mathrm{abs}$ of power absorbed by the membrane, since it can be shown that on cavity resonance ($\Delta_\mathrm{c}=0$) and for $\Ocnolambda^2\ll\kappa_\mathrm{c}\gamma$,
\begin{equation}
\label{eq:fabs}
f_\mathrm{abs}=\frac{4\Ocnolambda^2}{\gamma\kappa_\mathrm{c}}\,.
\end{equation}
It is worth noting the physical significance of \eref{eq:fabs}:\ in our model $\Ocnolambda$ and $\Oinnolambda$ effectively give rise to the imaginary part of the refractive index of the membrane, conventionally labelled $\kappa$. $f_\mathrm{abs}$ therefore exhibits a wavelength-dependence, as does $\kappa$, cf.~\tref{tab:FitParams}. Independent experimental measurements give us values for:\ $\kappa_\mathrm{c}$, $\Ocnolambda^2/\gamma$, $\omega_\mathrm{m}$, and $\tau_\mathrm{th}$. $g_0$ is fixed by the geometry and by the reflectivity of the membrane, whereas $P_\mathrm{in}$, $\omega_\mathrm{L}$, and $\Delta_\mathrm{c}$ are determined by the experiment. The only independent fit parameter in \eref{eq:kth} is therefore the photothermal coupling strength $\eta_\mathrm{th}/\gamma$.

\section{Fit to experimental data}\label{sec:Exp}
We shall now use $\kappa_\mathrm{th}$ to model four sets of experimental data. The experimental runs differ in the transverse location of the membrane vis-\`a-vis the cavity field, and the values of $\eta_\mathrm{th}$ obtained for the four sets are consistent with the coupling of the excitons to the $(2,1)$ drumhead mode of the membrane.\par
\begin{figure}[t]
 \centering
(a)\ \vtop{%
 \vskip-1ex
 \hbox{%
  \includegraphics[width=\figurewidth]{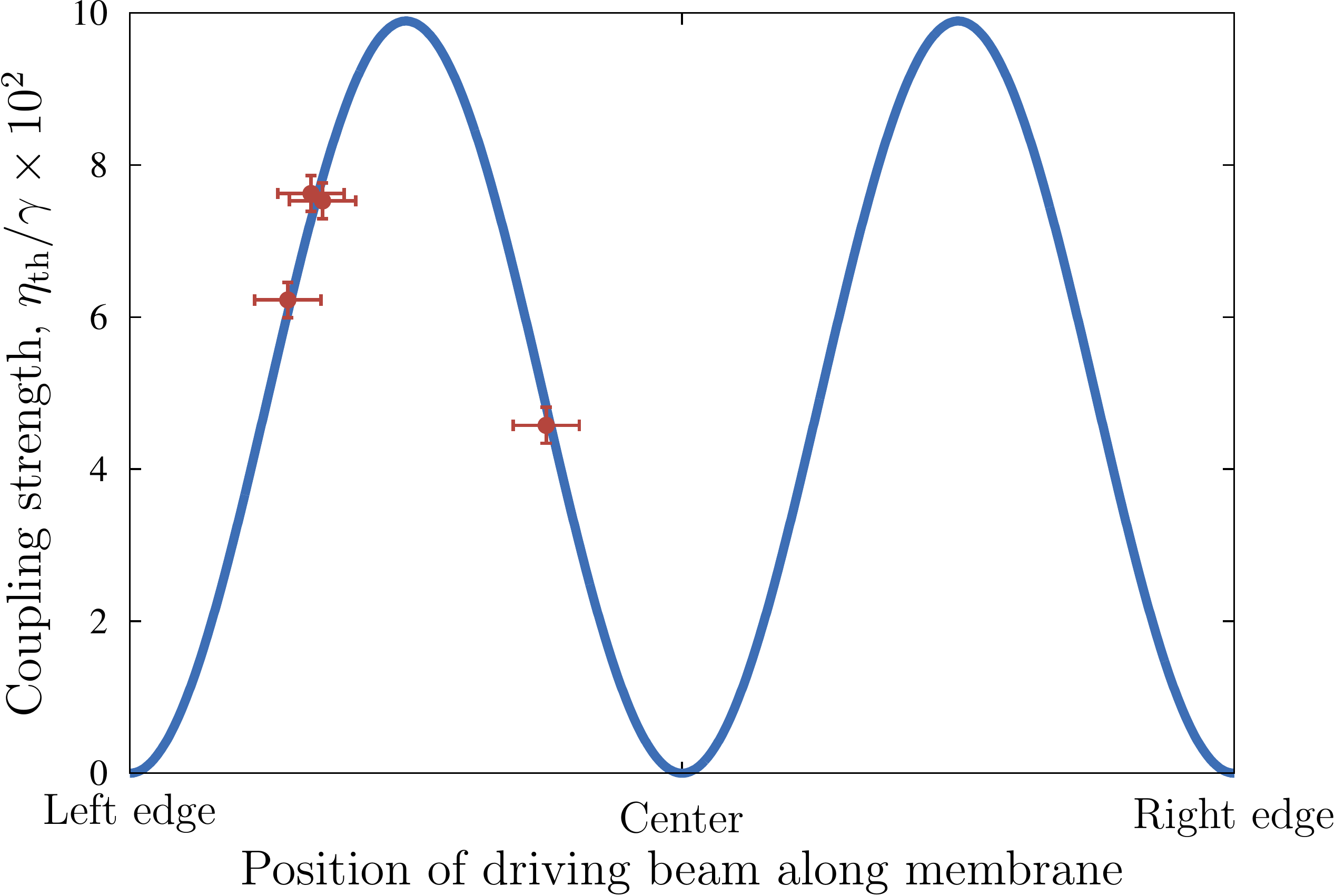}%
 }%
}\quad
(b)\ \vtop{%
 \vskip2ex
 \vskip-1ex
 \hbox{%
  \includegraphics[scale=0.5]{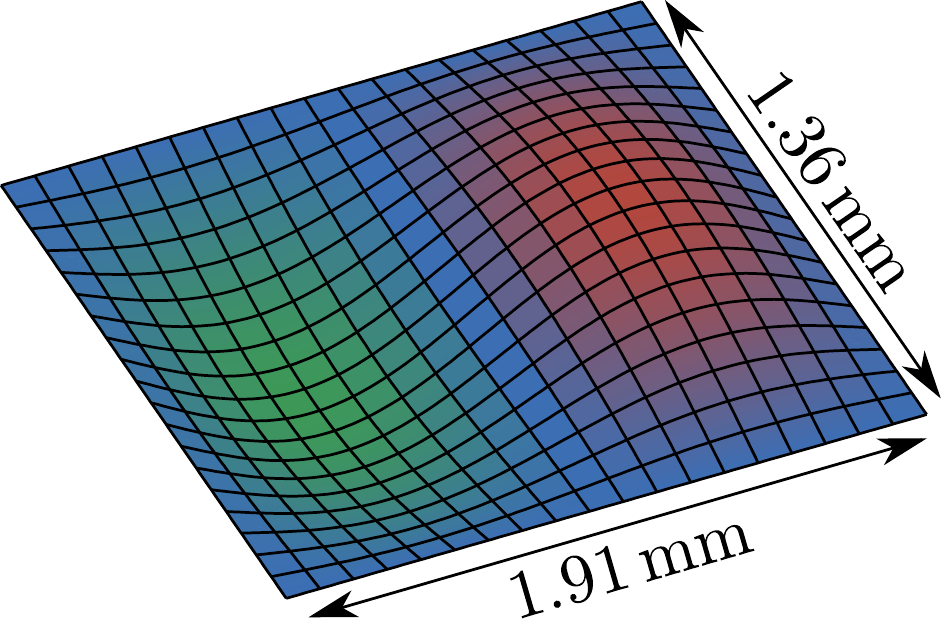}%
 }%
} \caption{(a) The coupling parameters that fit the four data sets are consistent with the excitation of the $(2,1)$ mode of the membrane [shown in (b)]; the solid curve represents the sinusoidal profile of the squared amplitude of this mode. (b) Vibrational profile of the $(2,1)$ mode of the membrane.}
 \label{fig:Couplings}
\end{figure}
The independently-determined parameters used in the model and common to every dataset were as follows:\ $L=2.9$\,cm, $d=160$\,nm, $g_0=2\pi\times(-5.1)$\,Hz, $\kappa_\mathrm{c}=2\pi\times258$\,MHz, $\omega_\mathrm{m}=2\pi\times23.4$\,kHz, and $\tau_\mathrm{th}=6.6$\,ms. Other parameters differed between datasets and are listed in \tref{tab:FitParams}. This table also lists the photothermal coupling rates resulting in the fits shown in \fref{fig:Fits}. These coupling rates, when plotted as a function of the position of the driving beam on the membrane, are consistent with a membrane displacement profile matching the $(2,1)$ drumhead mode, cf.\ \fref{fig:Couplings}, and therefore conform to our expectations. In other words, a single fit parameter, corresponding to a maximal coupling strength $\eta_\mathrm{th}^\mathrm{max}/\gamma=0.099$, suffices to fit \emph{all} the datasets if we also take into account the location of the driving beam on the membrane.

\section{Cooling through electronic stress}
Let us now consider a different system where the mechanism that provides the cooling force is no longer the thermal stress set up by decaying excitons but pure electronic stress. For a system such as the one we described above to enter this regime, the photothermal effect must be switched off. In the case of GaAs, this can be achieved by operating at a temperature of ca.\ $12$\,K~\cite{Usami2012}. Under these conditions, the excitons interact with the mechanics directly, and only through deformation potential, as expressed in the model of \sref{sec:HamEoM}. In an equivalent picture, we may say that the effective memory time is zero, and the memory kernel reduces to a delta-function:\ $\mathcal{M}(t)\to\delta(t)$.\\
By placing the membrane \emph{inside} a cavity~\cite{Thompson2008} ($\Oinnolambda=0$), the finesse of the cavity can be made significantly larger if level of absorption in the membrane is lowered, leading to a correspondingly smaller $\Ocnolambda$. This can be done in GaAs by running the experiment at longer wavelengths, \eg, at $884$\,nm, where the absorption is significantly lower than at $870$\,nm.

\section{Conclusion}
We have explored optomechanical cooling through the photothermal effect in a semiconductor membrane. Our model uses a coupling similar to the deformation-potential coupling but makes use of a memory kernel to model the long thermalisation time typical of such structures. The introduction of the memory kernel was based on entirely phenomenological grounds, following Refs.~\cite{Metzger2008,Pinard2008,DeLiberato2011,Restrepo2011}, but we justify the use of an exponentially-decaying kernel by introducing additional phononic degrees of freedom that are then eliminated. The resulting model only has one free parameter, with all others being determined independently or by the geometry of the situation, and provides a remarkably good fit to the experimental data.\par
By using different forms of the memory kernel we can also compare the different physical mechanisms in promoting optomechanical cooling. Thus, for example, an instantaneous memory kernel $\mathcal{M}(t)=\delta(t)$ reduces our description to one taking into account pure deformation-potential effects, similarly to what was originally envisioned in Ref.~\cite{WilsonRae2004}. Such a mechanism could be important under conditions where the photothermal effect is cancelled out, \eg, at temperatures where the membrane undergoes no photothermal deformation~\cite{Usami2012}.

\section*{Acknowledgements}
KH and AX acknowledge support through the Centre for Quantum Engineering and Space-Time Research (QUEST) at the Leibniz University Hannover. AX also acknowledges support from the Royal Commission for the Exhibition of 1851. We thank Ata\c{c} \.{I}mamo\={g}lu and Ignacio Wilson-Rae for illuminating discussions.

\appendix
\section{Indirect exciton--motion coupling, and emergence of memory kernel}\label{app:MemoryKernel}
The model presented in \sref{sec:HamEoM} serves to explain the physical processes occurring in a more transparent manner. As far as the interaction of the $\hat{c}_\lambda$ with $\hat{b}$ is concerned, however, this model does not capture the fact that the process takes place indirectly. In this Appendix we will examine a more detailed Hamiltonian that leads to the same effective equation of motion for $\hat{b}$. We shall introduce a phononic bath of modes $\hat{d}_\mu$ (oscillation frequency $\omega_\mu$ and amplitude decay rate $\kappa_\mu$) that serve as intermediaries between $\hat{c}_\lambda$ and $\hat{b}$. Physically, the $\hat{d}_\mu$ account for the nonzero temperature of the lattice making up the membrane. The interaction terms between the three systems can be written as~\cite{Toyozawa1958,SteynRoss1983}
\begin{equation}
\label{eq:BathHamiltonianTerms}
\sum_{\lambda,\lambda^\prime,\mu}k_{\lambda,\lambda^\prime,\mu}\hat{c}_\lambda^\dagger\hat{c}_{\lambda^\prime}\bigl(\hat{d}_\mu+\hat{d}_\mu^\dagger\bigr)+\sum_{\mu,\mu^\prime}l_{\mu,\mu^\prime}\bigl(\hat{d}_\mu^\dagger\hat{d}_{\mu^\prime}\hat{b}+\hat{d}_\mu\hat{d}_{\mu^\prime}^\dagger\hat{b}^\dagger\bigr)\,,
\end{equation}
with the $k_{\lambda,\lambda^\prime,\mu}$ and $l_{\mu,\mu^\prime}$ being coupling frequencies whose values we shall not specify or calculate. Hermiticity requires that $k_{\lambda^\prime,\lambda,\mu}=k_{\lambda,\lambda^\prime,\mu}^\ast$ and $l_{\mu^\prime,\mu}=l_{\mu,\mu^\prime}^\ast$. Any sums over $\lambda$ or $\mu$ may be either discrete or continuous, as the case requires. \eref{eq:BathHamiltonianTerms} mediates the interaction between the excitons and $\hat{b}$, and therefore replaces the second term in $\hat{H}_\mathrm{mech}$.\\
We shall now proceed to eliminate the phononic bath modes. The equation of motion for the $\hat{d}_\mu$ reads
\begin{equation}
\label{eq:EoMdmu}
\dot{\hat{d}}_\mu=-\bigl(\kappa_\mu+i\omega_\mu\bigr)\hat{d}_\mu-i\sum_{\lambda,\lambda^\prime}k_{\lambda,\lambda^\prime,\mu}\hat{c}_\lambda^\dagger\hat{c}_{\lambda^\prime}-i\sum_{\mu^\prime}\bigl(l_{\mu,\mu^\prime}\hat{d}_{\mu^\prime}\hat{b}+l_{\mu^\prime,\mu}\hat{d}_{\mu^\prime}\hat{b}^\dagger\bigr)\,,
\end{equation}
where we have not written down the input noise terms, expressed in terms of the anti-Hermitian operators $\hat{d}_{\mu,\mathrm{in}}$ in the quantum Brownian-motion damping model~\cite{Giovannetti2001}, since these terms have no effect on the cooling \emph{rate} but help to determine the lowest mechanical occupation number that can be achieved through this cooling mechanism.\\
The last set of terms in \eref{eq:EoMdmu} leads, both directly and through the $\hat{c}_\lambda$, to a renormalisation of $\omega_\mathrm{m}$ and $\kappa_\mathrm{m}$ due to the absorbed optical power and finite temperature of the lattice, and we may therefore safely ignore it, linearise the equation of motion, and finally write
\begin{equation}
\dot{\hat{d}}_\mu=-\bigl(\kappa_\mu+i\omega_\mu\bigr)\hat{d}_\mu-i\sum_{\lambda,\lambda^\prime}\bigl(k_{\lambda,\lambda^\prime,\mu}\bar{c}_\lambda^\star\hat{c}_{\lambda^\prime}+k_{\lambda^\prime,\lambda,\mu}\bar{c}_\lambda\hat{c}_{\lambda^\prime}^\dagger\bigr)\,.
\end{equation}
Formally, then, the solution for $\hat{d}_\mu$ is given by
\begin{equation}
\hat{d}_\mu(t)=-i\int_{-\infty}^te^{-(\kappa_\mu+i\omega_\mu)(t-\tau)}\sum_{\lambda,\lambda^\prime}\bigl[k_{\lambda^\prime,\lambda,\mu}\bar{c}_{\lambda^\prime}^\star\hat{c}_\lambda(\tau)+k_{\lambda,\lambda^\prime,\mu}\bar{c}_{\lambda^\prime}\hat{c}_\lambda^\dagger(\tau)\bigr]\rmd\tau\,,
\end{equation}
noting once more that we are ignoring input noise fields. The Hamiltonian above therefore gives the following contribution to the linearised equation of motion for $\hat{b}$:
\begin{eqnarray}
\fl\dot{\hat{b}}&=-i\sum_{\mu,\mu^\prime}\bigl(l_{\mu,\mu^\prime}\bar{d}_{\mu^\prime}^\star\hat{d}_\mu+l_{\mu^\prime,\mu}\bar{d}_{\mu^\prime}\hat{d}_\mu^\dagger\bigr)\nonumber\\
\fl&=-i\int_{-\infty}^t\sum e^{-\kappa_\mu(t-\tau)}\Bigl\{-ie^{-i\omega_\mu(t-\tau)}l_{\mu,\mu^\prime}\bar{d}_{\mu^\prime}^\star\bigl[k_{\lambda^\prime,\lambda,\mu}\bar{c}_{\lambda^\prime}^\star\hat{c}_\lambda(\tau)+k_{\lambda,\lambda^\prime,\mu}\bar{c}_{\lambda^\prime}\hat{c}_\lambda^\dagger(\tau)\bigr]\Bigr\}\rmd\tau\nonumber\\
\fl&\qquad-i\int_{-\infty}^t\sum e^{-\kappa_\mu(t-\tau)}\Bigl\{ie^{i\omega_\mu(t-\tau)}l_{\mu^\prime,\mu}\bar{d}_{\mu^\prime}\bigl[k_{\lambda,\lambda^\prime,\mu}^\star\bar{c}_{\lambda^\prime}^\star\hat{c}_\lambda(\tau)+k_{\lambda^\prime,\lambda,\mu}^\star\bar{c}_{\lambda^\prime}\hat{c}_\lambda^\dagger(\tau)\bigr]\Bigr\}\rmd\tau\nonumber\\
\fl&=-i\int_{-\infty}^t\sum e^{-\kappa_\mu(t-\tau)}\bigl[-ie^{-i\omega_\mu(t-\tau)}l_{\mu,\mu^\prime}\bar{d}_{\mu^\prime}^\star+\mathrm{c.c.}\bigr]k_{\lambda^\prime,\lambda,\mu}\bar{c}_{\lambda^\prime}^\star\hat{c}_\lambda(\tau)\rmd\tau\nonumber\\
\fl&\qquad-i\int_{-\infty}^t\sum e^{-\kappa_\mu(t-\tau)}\bigl[-ie^{-i\omega_\mu(t-\tau)}l_{\mu,\mu^\prime}\bar{d}_{\mu^\prime}^\star+\mathrm{c.c.}\bigr]k_{\lambda,\lambda^\prime,\mu}\bar{c}_{\lambda^\prime}\hat{c}_\lambda^\dagger(\tau)\rmd\tau\,,
\end{eqnarray}
with the sums running over $\lambda$, $\lambda^\prime$, $\mu$, and $\mu^\prime$; `$\mathrm{c.c.}$' denotes the complex conjugate of the preceding term. We now make the formal replacement
\begin{equation}
\fl\sum_{\lambda^\prime,\mu,\mu^\prime}e^{-\kappa_\mu(t-\tau)}\bigl[-ie^{-i\omega_\mu(t-\tau)}l_{\mu,\mu^\prime}\bar{d}_{\mu^\prime}^\star+\mathrm{c.c.}\bigr]k_{\lambda^\prime,\lambda,\mu}\bar{c}_{\lambda^\prime}^\star\to\eta^\star\Bigl(\sum_\lambda\bar{c}_\lambda^\star\Bigr)\,\mathcal{M}(t-\tau)\,,
\end{equation}
where $\mathcal{M}(t)$ is a causal memory kernel that we choose to have a decaying exponential form, and where $k_{\lambda^\prime,\lambda,\mu}$ is assumed to be independent of $\lambda$ and $\lambda^\prime$. Finally, then, the contribution to the equation of motion for $\hat{b}$ is
\begin{eqnarray}
\dot{\hat{b}}=&-i\eta^\star\Bigl(\sum_\lambda\bar{c}_\lambda^\star\Bigr)\int_{-\infty}^\infty\mathcal{M}(t-\tau)\sum_\lambda\hat{c}_\lambda(\tau)\rmd\tau\nonumber\\
&\qquad-i\eta\Bigl(\sum_\lambda\bar{c}_\lambda\Bigr)\int_{-\infty}^\infty\mathcal{M}(t-\tau)\sum_\lambda\hat{c}_\lambda^\dagger(\tau)\rmd\tau\,.
\end{eqnarray}
The memory kernel in the equation of motion for $\hat{b}$ therefore arises naturally from this more complete, albeit still phenomenological, model.

\section*{References}
\bibliographystyle{iopart-num}
\providecommand{\newblock}{}

\end{document}